\documentclass[aps,prd,twocolumn,groupedaddress]{revtex4-1}
\usepackage{amsmath}
\usepackage{graphicx}
\usepackage{dcolumn}

\begin{document}


\title{Forecasting isocurvature models with CMB lensing information: axion and curvaton scenarios}


\author{L. Santos}
\email{larissa.santos@roma2.infn.it}
\affiliation{Dipartimento di Fisica, Universit\`a di Roma ``Tor Vergata''}
\author{P. Cabella}
\affiliation{Dipartimento di Fisica, Universit\`a di Roma ``Tor Vergata''}
\author{A. Balbi and N. Vittorio}
\affiliation{Dipartimento di Fisica, Universit\`a di Roma ``Tor Vergata"}
\affiliation{INFN, Sezione di Roma Tor Vergata}


\date{\today}

\begin{abstract}
Some inflationary models predict the existence of isocurvature primordial fluctuations, in addition to the well known adiabatic perturbation. Such mixed models are not yet ruled out by available data sets. In this paper we explore the possibility of obtaining better constraints on the isocurvature contribution from future astronomical data. We consider the axion and curvaton inflationary scenarios, and use Planck satellite experimental specifications together with SDSS galaxy survey to forecast for the best parameter error estimation by means of the Fisher information matrix formalism. In particular, we consider how CMB lensing information can improve this forecast. We found substantial improvements for all the considered cosmological parameters. In the case of isocurvature amplitude this improvement is strongly model dependent, varying between less than 1\% and above 20\% around its fiducial value. Furthermore, CMB lensing enables the degeneracy break between the isocurvature amplitude and correlation phase in one of the models. In this sense, CMB lensing information will be crucial in the analysis of future data.
\end{abstract}


\maketitle

\section{Introduction\label{Introduction}}

Since the early measurement of the first acoustic peak in the Cosmic Microwave Background (CMB) angular power spectrum \citep{2000debernardis, hanany2000}, a pure isocurvature model of primordial fluctuations was ruled out \citep{2000enqvist}.  In addition,  recent CMB data from the WMAP satellite found no evidence for non-adiabatic primordial fluctuations \citep{2011komatsu}. These results are consistent with a single scalar field inflationary model prediction of  perfectly adiabatic density perturbations. However, small contributions from isocurvature primordial fluctuations (in mixed models) cannot be excluded by the current data.

The standard inflationary scenario driven by a single field  cannot account for isocurvature fluctuations. If we want to take into account isocurvature fluctuations, a multiple-field inflation has to be considered (for a general formalism, see \citet{2001gordon} for example). We consider the alternative scenario where perturbations to a light field different from the inflaton (the curvaton) are responsible for curvature perturbations and may also generate isocurvature fluctuations \citep{2002lyth, 2002moroi, 2003lyth}. In this case, the isocurvature component is completely correlated or anti-correlated with the adiabatic component.  A second scenario is also taken into account in this work, where quantum fluctuations in a light axion field generate isocurvature fluctuations.  Unlike the first scenario, this isocurvature component is fully uncorrelated with the adiabatic one. It is important to point out that axion particles can be produced in this scenario, which can contribute to the present dark matter in the universe (see \citet{2007beltran, 2002bozza, 2008hertzberg} and references therein).

There are  already many studies in the literature constraining the isocurvature contribution using different data sets (CMB, large-scale structure (LSS), type Ia supernovae (SN), Lyman-$\alpha$ forest and baryon acoustic oscillations (BAO)) (see, for example, \citet{2011li, 2011carbone, 2011larson, 2011komatsu, 2010mangilli, 2006bean, 2005beltran, 2003crotty}). We intend to use the well-known Fisher information matrix formalism (for a short guide see \citet{2009coe}) to estimate whether better constrains to the isocurvature contribution can be obtained in the near future using measurements of the CMB temperature and polarization power spectrum from the Planck satellite, as well as the large-scale matter distribution observed by the Sloan Digital Sky Survey (SDSS), using CMB lensing information. We will see how this new information could improve the error prediction for some cosmological parameters, especially those related to the isocurvature mode.

The paper is organized as follows: in Section \ref{Isocurvature notation} we describe briefly the isocurvature models and the notation that will be used throughout the paper. We give a small introduction on CMB lensing in Section \ref{CMB lensing}. In Section \ref{Method},  we briefly review the Fisher information matrix formalism for the CMB (with and without lensing information) and for a galaxy survey. Finally, we present our results in Section \ref{Results}, followed by our discussion and conclusions in Section \ref{Discussion and conclusions}.

\section{Isocurvature notation\label{Isocurvature notation}}

In this paper, we consider the standard isocurvature Cold Dark Matter (CDM) mode generated during inflationary time (For a more general case that also consider other generated modes, as for example the baryon mode, see \citet{2000bucher}), that reads:

\begin{equation}
\label {iso_mode}
S_{CDM} = \frac{\delta \rho_{CDM} }{ \rho_{CDM} } -\frac{3}{4} \frac{\delta \rho_{\gamma} }{ \rho_{\gamma} }. 
\end{equation}

The adiabatic density perturbation, believed to be the major responsible for the formation of the observable structure in the universe defined as:

\begin{equation}
\label {ad_mode}
 \frac{1}{4}\frac{\delta \rho_{\gamma} }{ \rho_{\gamma} } =\frac{1}{4} \frac{\delta \rho_{\nu} }{ \rho_{\nu} } = \frac{1}{3} \frac{\delta \rho_{B} }{ \rho_{B} } = \frac{1}{3} \frac{\delta \rho_{CDM} }{ \rho_{CDM} },
\end{equation}
for a universe that consists of photons, massless neutrinos, baryons and CDM at early times. Following \citet{2006bean}, we write the isocurvature contribution to the purely adiabatic CMB power spectra in the form:

\begin{equation}
\label {iso_cmb}
\small
C_{l} = (1-\alpha)C_{l}^{ad} + \alpha C_{l}^{iso} +2\beta \sqrt{\alpha(1-\alpha)}C_{l}^{cross},
\end{equation}
\normalsize
where the parameter $\alpha$ accounts for the isocurvature amplitude, while $\beta$ stands for the isocurvature correlation phase, given by $\beta=\cos\theta$,  $-1 \leq \cos\theta \leq 1$.  In the same way, the matter power spectrum, $P(k)$, can be decomposed into purely adiabatic, purely isocurvature and their cross correlation contribution:

\begin{equation}
\label {iso_gal}
\small
P(k) = (1-\alpha) P(k)^{ad} + \alpha P(k)^{iso} +2\beta \sqrt{\alpha(1-\alpha)}P(k)^{cross},
\end{equation}
\normalsize
 where
 
\begin{equation}
\label {matter}
P^{i}(k) = T^{i}(k)\left( \frac{k}{k_0}\right)^{n_{i}-1}, i= \textrm{ad, iso or cross}
\end{equation}

As stated by \citet{2006bean} it is reasonable to assume a cross spectrum independent of scale, $n_{cross}= \frac{1}{2}(n_{ad} +n_{iso})$. We take the pivot value $k_0=0.002 Mpc^{-1}$ as used by the WMAP team.

\section{CMB lensing\label{CMB lensing}}

As new experiments are developed, the precision in CMB measurements  makes small effects distinguishable by observations. CMB lensing is one of this effects and it has important quantitative contribution that should be taken into account.  It is known that the CMB photons are deflected during their travel between the last scattering surface and the observer by gravitational potentials $\Psi (\chi, \eta)$ dependent on the comoving distance $\chi$ and the conformal time $\eta$. For CMB temperature anisotropy, this is quantitatively written as: 

\begin{equation}
\label {temp_lens}
\frac{\Delta\textrm{\~T(\textbf{\^n}})} {T} = \frac{\Delta\textrm{T(\textbf{\^n'})}}{T} = \frac{\Delta\textrm{T(\textbf{\^n}} + d)}{T},
\end{equation}
where the temperature T of the lensed CMB in a direction \textbf{\^n} is equal to the unlensed CMB in a different direction \textbf{\^n'}.  Both these directions, \textbf{\^n} and \textbf{\^n'}, differ by the deflection angle $d$ as it can be seen in the third equality above.  To first order, the deflection angle is simply the lensing potential gradient, $d= \nabla \psi$. In the same way,  the effect of lensing in CMB polarization is written in terms of the Stokes parameters $Q(\textbf{\^n})$ and $U(\textbf{\^n})$ (for a review in CMB polarization theory, see \citet{2004cabella}) :
\begin{equation}
\label {pol_lens}
[Q + iU] (\textbf{\^n})  = [Q+ iU] (\textbf{\^n}+ d).
\end{equation}

To use the CMB lensing information we the have to measure the lensing potential that is defined as:

\begin{equation}
\label {pot_lens}
\psi(\textbf{\^n}) \equiv -2\int_{0}^{\chi^*}d\chi \frac{\chi^*- \chi} {\chi^*\chi} \Psi(\chi\textbf{\^n}; \eta_{0}-\chi),
\end{equation}
being $\chi^*$ the comoving distance and $\eta_{0}-\chi$ is the conformal time at which the photon was at position $\chi\textbf{\^n}$.

We can study CMB lensing properties through the lensing potential, and the temperature and polarization power spectra, as well as their cross-correlation  (for a review of CMB lensing see \citet{2006lewis}).  

The lensing signal was detected for the first time by cross-correlating WMAP data to radio galaxy counts in the NRAO VLA sky survey (NVSS) \citep{2007smith}. In other words, $C^{\psi g} \neq 0$: however, it is still not possible to obtain the lensing potential power spectrum, $C^{\psi \psi}$, from current data.  While waiting for more precise data sets, much work is being done to CMB lensing reconstruction techniques (e.g. \citet{2001hu, 2003okamoto, 2010smith, 2010bucher, 2011carvalho}). In this paper, we use the CAMB software package   (http://camb.info) \citep{2000lewis} to obtain the numerical lensed and unlensed power spectra  ($C^{TT},  C^{EE}, C^{BB}, C^{TE}$ and $C^{dd}, C^{T d}$) for each cosmological model. We then used this predictions to forecast how CMB lensing information will help us constraining some isocurvature models when more precise future experiment will be available and the lensing potential can be extracted from data.

\section{Method\label{Method}}

In order to search how an isocurvature contribution would affect the measurements of cosmological parameters we apply the Fisher information matrix formalism to a Planck-like experiment \citep{2005bb}, considering both temperature and polarization for the lensend and unlensed CMB spectrum,  and to the Sloan Digital Sky Survey (SDSS).  We consider both the axion and the curvaton scenarios in a $\Lambda$CDM model. 

\subsection{Information from CMB}

The Fisher information matrix for the CMB temperature anisotropy and polarization is given by  the approximation in \citep{1997zaldarriaga}

\begin{equation}
\label {fisher_unl}
F_{ij} =\sum_l\sum_{XY}  \frac{\partial C^{X}_l} {\partial p_i} (Cov_l^{-1})_{XY}  \frac{\partial C^{Y}_l} {\partial p_j},
\end{equation}
where $C^{X}_l $ is the power in the $l$th multipole, $X$ stands for $TT$ (temperature), $EE$ (E-mode  polarization), $BB$ (B-mode polarization) and $TE$ (temperature and E-mode polarization cross-correlation). We will not include primordial B-modes in the analysis since the measurement of the primordial $ C^{BB}_l$ by Planck is expected to be noise dominated. Our covariance matrix becomes therefore:

\small
\begin{equation}
\label {cov_array_unl}
Cov_l= \frac {2} {(2l+1)fsky} \left[\begin{array}{rrr}
\Xi_{TTTT} & \Xi_{TTEE} & \Xi_{TTTE}\\
\Xi_{EETT} & \Xi_{EEEE} & \Xi_{EETE}\\
\Xi_{TETT} & \Xi_{TEEE} & \Xi_{TETE}
\end{array}\right].
\end{equation}
\normalsize
Explicit expressions for the matrix elements are given in the appendix.

For the lensed case we have to perform a correction in the covariance matrix elements taking into consideration the power spectrum of the deflection angle and its cross correlation with temperature, $C^{Td}_l$. We also change in this case the unlensed CMB power spectra, $C^{X}_l$, for the lensed ones, $\textrm{\~C}^{X}_l$. 
When we include these corrections, the covariance matrix becomes:

\small
\begin{equation}
\label {cov_array_lens}
\begin{split}
& Cov_l=  \frac {2} {(2l+1)fsky} \\
& \left[\begin{array}{rrrrrr}
\xi_{TTTT} & \xi_{TTEE} & \xi_{TTTE}      & \xi_{TTTd}     & \xi_{TTdd} & 0 \\
\xi_{EETT} & \xi_{EEEE} & \xi_{EETE}    & 0                     & 0                 &0  \\
\xi_{TETT} & \xi_{TEEE} & \xi_{TETE}     & 0                     &0                   &0 \\
\xi_{TdTT} & 0                    &0                     &  \xi_{TdTd}    & \xi_{Tddd}  &0 \\
\xi_{ddTT} & 0                    &0                     &  \xi_{ddTd}    & \xi_{dddd}  &0 \\
               0  & 0                    &0                     &0                       &0                  &\xi_{BBBB}
\end{array}\right].
\end{split}
\end{equation}
\normalsize

Full expressions for the corrections to the covariance matrix can be found in the appendix. Note that in this case we are taking into consideration the B-mode polarization generated by the CMB gravitational lensing from the E-mode polarization. In both cases, we used $fsky=0.65$.

\subsection{Information from galaxy survey}

The Fisher information matrix for the matter power spectrum obtained from galaxy surveys is given by \citep{1997tegmark}:

\begin{equation}
\label {fisher_gal}
F_{ij} =\int_{k_{min}}^{k_{max}} \frac{\partial \ln P(k)} {\partial p_i}  \frac{\partial \ln P(k)} {\partial p_j}V_{eff} \frac{k^2 dk} {(2\pi)^2}  ,
\end{equation}

\begin{equation}
\label {veff}
V_{eff}(k) =\int \left[ \frac{\bar{n}(r) P_g(k)} {1+\bar{n}(r)P_g(k)} \right]^2 d^3r .
\end{equation}

We know that $P_g(k)=b^2P(k)$ and using the specifications of SDSS experiment for the Bright Red Galaxy (BRG) sample, called Luminous Red Galaxies (LRG) in more recent papers, we assume a linear and scalar independent bias $b=2$ \citep{1997tegmark,2006hutsi}.  It is assumed that the expected number density of galaxies, $\bar{n}(r)$, is independent of $r$, $n=10^5 /V_s$, in a volume-limited sample to a depth of $10^3$Mpc. The survey has an angle of $\pi$ steradians, therefore the survey volume becomes, $V_s= 10^9 \pi/3$ \citep{1997tegmark, 1999eisenstein, 2001eisenstein}. 

\section{Results\label{Results}}

For the purpose of our analysis, we consider as free cosmological parameters the adimensional value of the Hubble constant, $h$, the density of baryons and cold dark matter, $\Omega_b h^2$ and $\Omega_c h^2$, the spectral index of scalar adiabatic perturbations,  $n_{ad}$, the aforementioned isocurvature parameters, $\alpha$ and $\beta$, and the equation of state of dark energy $w$, assumed as a constant.  We first performed the forecast for Planck alone, with and without considering CMB lensing. Then, we introduced the forecast for SDSS, combining the results as (assuming that SDSS and CMB results can be well approximated as independent ones):

\begin{equation}
\label {fisher_unl_len_sdssl}
F^{Total}_{ij} = F^{Planck}_{ij} + F^{SDSS}_{ij}.
\end{equation}

We considered 3 different scenarios to constrain the cosmological parameters.  

First we use an axion scenario considering that a high $n_{iso}=1.9 \pm 1$ is favored by Lyman-$\alpha$ data \citep{2005beltran}.  \citet{2009kasuya} proposed an axion model capable of generating isocurvature fluctuations with an extremely blue spectrum with $1 < n_{iso} \leq 4$. Taking into consideration that Planck could measure the existence of isocurvature contribution with a high spectral index motivates our parameter forecast of such model.  We choose a fixed $n_{iso}=2.7$ considering that \citet{2005beltran} tested the robustness of their result finding an extreme model with $n_{iso} = 2.7$. \citet{2006bean} found this same value for the isocurvature spectral index for their best fit model considering an adiabatic plus isocuvarture CDM contribution, however for a generally correlated isocurvature component with respect to the adiabatic one. It was shown, however, that the chosen pivot scale affects the $n_{iso}$ likelihood \citep{2005Kurki-Suonio}. The previous mentioned articles  \citep{2005beltran,2006bean} used a pivot scale $k_0 = 0.05 Mpc^{-1}$ that favors artificially large $n_{iso}$ according to \citet{2005Kurki-Suonio}. They chose instead $k_0 =0.01 Mpc^{-1}$ and found that the likelihood for  $n_{iso}$ peaks at approximately 3.  It was also shown that for $k_0 < 0.01 Mpc^{-1}$ the results does not change drastically, concluding that our choice for $n_{iso}=2.7$ is valid. 

\begin{figure*} [h!]
\includegraphics[scale=0.3]{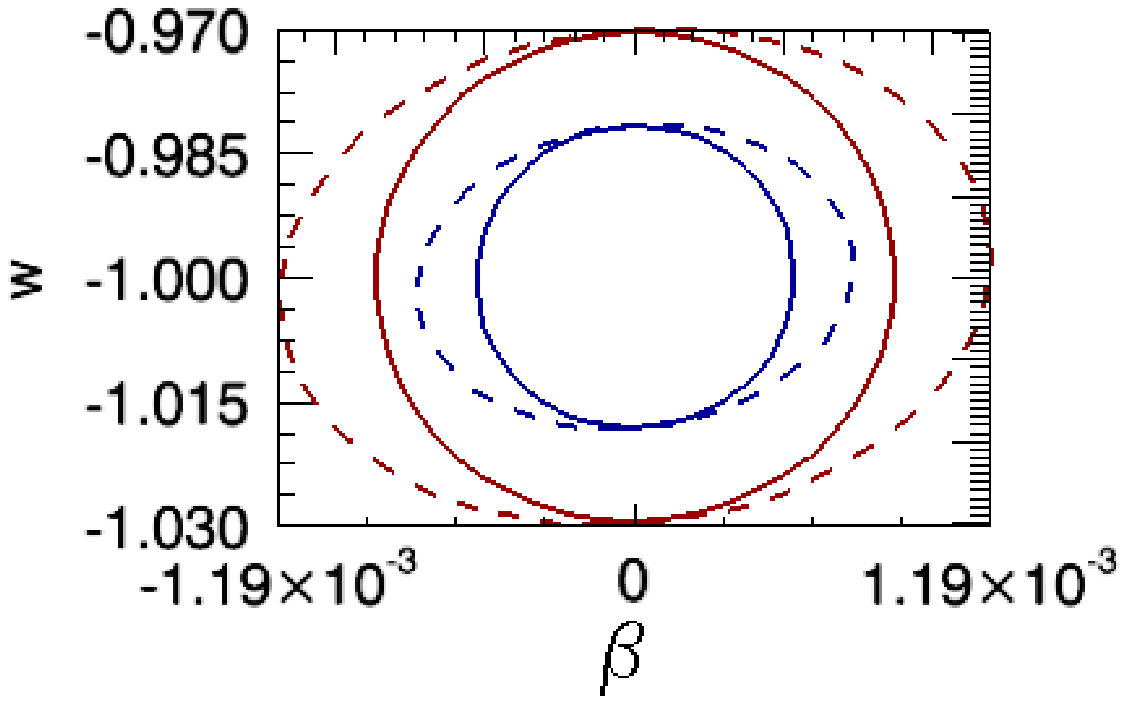}
\includegraphics[scale=0.3]{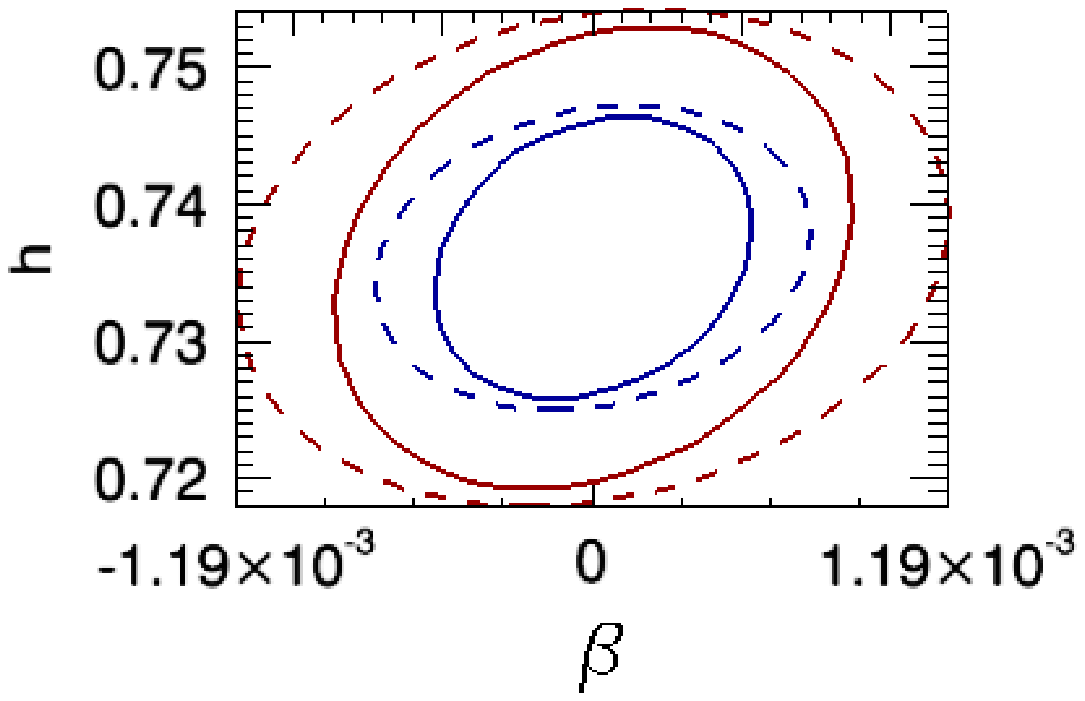}
\includegraphics[scale=0.3]{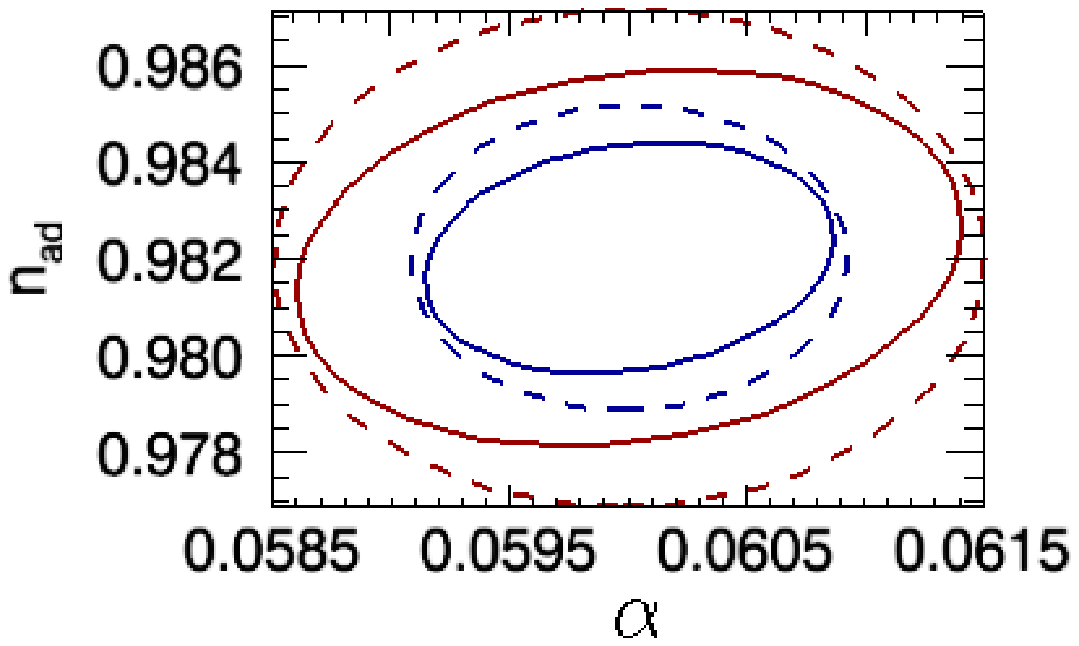} \\

\includegraphics[scale=0.3]{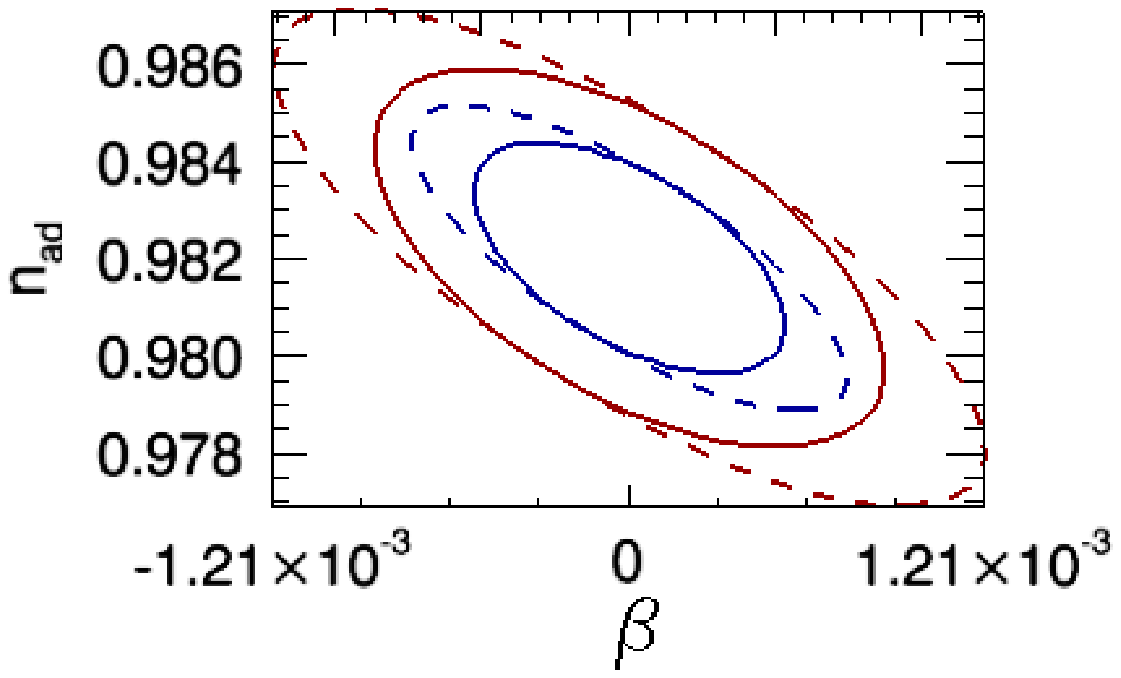} 
\includegraphics[scale=0.3]{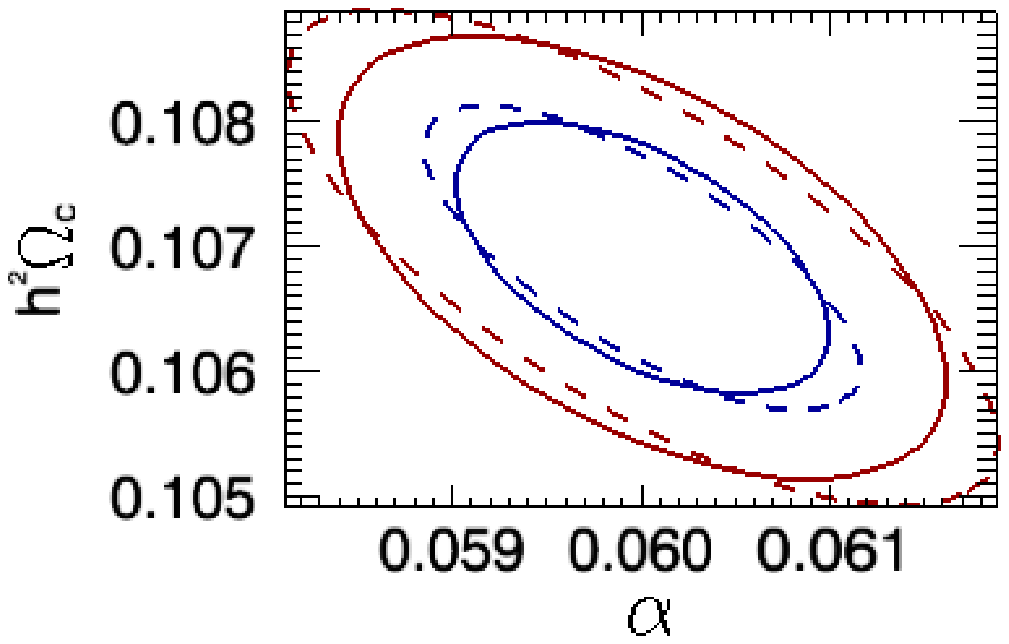}
\includegraphics[scale=0.3]{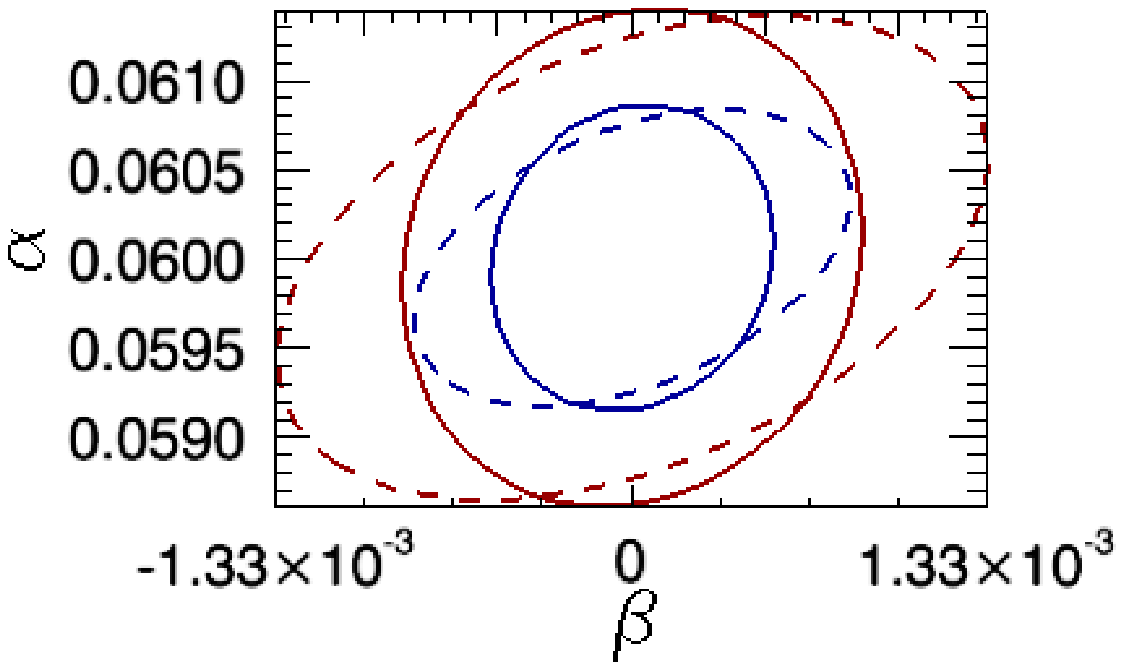} 
\includegraphics[scale=0.3]{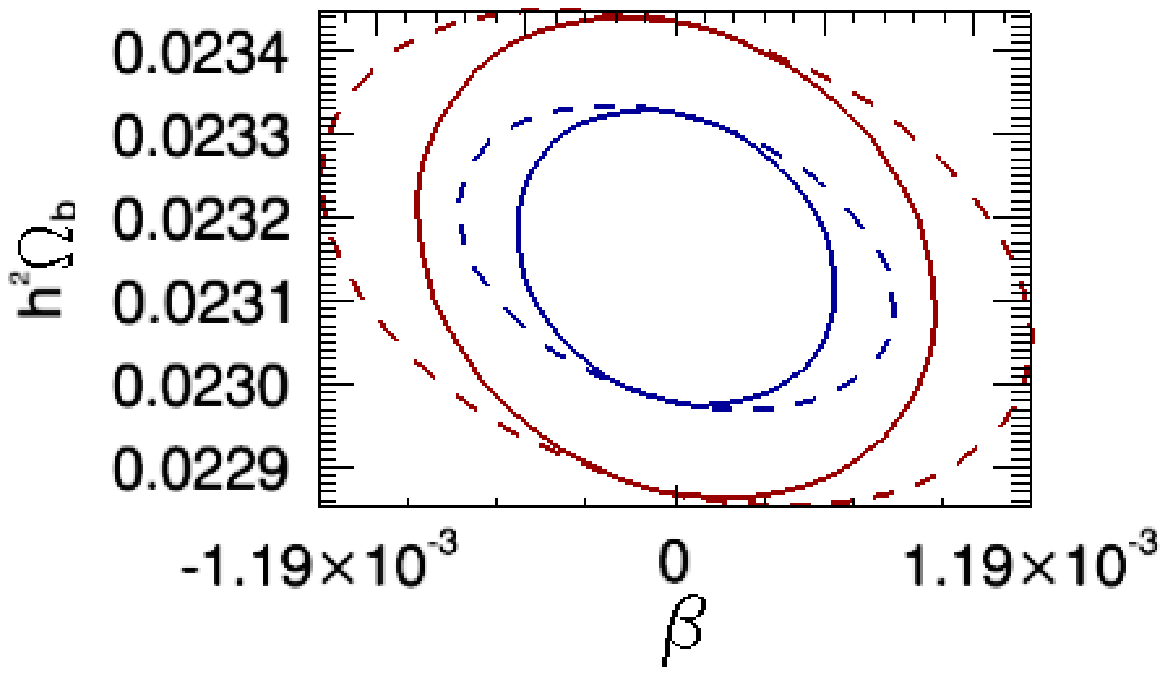} 

\caption{Fisher contours for  $\Lambda$CDM model plus a contribution of initial isocurvature fluctuation with fiducials amplitude of $\alpha = 0.06$, correlation phase of $\beta = 0$ and scalar spectral index of $n_{iso} = 2.7$ for the axion scenario. The red and blue contours represent 95.4\% and 68\% C.L. respectively for the unlesed CMB + SDSS (dashed lines) and for the lensed CMB + SDSS (solid lines)(see Table \ref {tbl-modelo1}). } 
\label{fisher_modelo1}
\end{figure*}

In this case our fiducial model is given by $h=0.736$, $\Omega_b h^2=0.02315$, $\Omega_c h^2=0.1069$, $n_{ad} = 0.982$  ($n_{iso} = 2.7$ fixed) $\alpha =0.06$, $\beta= 0$ and $w=-1$. Fisher contours and all the 1 sigma errors are shown in Figure \ref{fisher_modelo1} and Table \ref{tbl-modelo1}. Our first approach was to let $\beta$ vary, obtaining in this way a lower and upper limit to the isocurvature correlation phase. Even if not predicted by the chosen inflationary scenario (axion type), we can still constrain a possibly non-zero measurement of $\beta$ where this scenario is still valid.  A second approach was to keep $\beta$ fixed, as it is showed in Table \ref{tbl-modelo1_sem_beta}. It can be noticed that there is not a significant change in the constraints on the other cosmological parameters between Tables \ref {tbl-modelo1}  and \ref{tbl-modelo1_sem_beta},  especially for $\alpha$.

\begin{table*}
\begin{center}
\caption{Marginalized errors for $\Lambda$CDM model plus a contribution of initial isocurvature fluctuation with fiducials amplitude of $\alpha = 0.06$, correlation phase of $\beta = 0$ and scalar spectral index of $n_{iso} = 2.7$.\label{tbl-modelo1} }
\begin{tabular}{cccccc}
\tableline
Parameter & CMB alone & CMB alone & P(k) alone &CMB + P(k)          & CMB + P(k)     \\
                    & Planck         & Planck         & SDSS        & Planck + SDSS   & Planck +SDSS \\
                    & T + P            & T+ P+ lens  &                    & T + P                     & T + P + lens       \\
 \hline \hline

 $h$                           &0.031     &0.021         &0.43    &0.0073        & 0.0068    \\ 
 $h^2 \Omega_b$  &0.00013 &0.00012    &0.038  &0.00012     &0.00012   \\ 
 $h^2 \Omega_c$  &0.0011   &0.00089    &0.14     & 0.00093    &0.00079   \\
 $n_s$                      &0.0081   &0.0045      & 0.70   &0.0021        &0.0016     \\
 $\alpha$                 &0.0011   &0.00089   &0.28     &0.00060     &0.00056  \\
 $\beta$                   &0.00077 &0.00040   &0.17    &0.00048     &0.00035   \\
 w                             &0.071      &0.048        &2.21    &0.012         &0.012       \\
 \hline
 \end{tabular} 

 \begin{tabular}{c}
          Percentage of the parameters' fiducial values for each error above  \\
   \end{tabular} 
   
  \begin{tabular}{cccccc}
  \hline
Parameter & CMB alone & CMB alone & P(k) alone &CMB + P(k)          & CMB + P(k)     \\
                    & Planck         & Planck         & SDSS        & Planck + SDSS   & Planck +SDSS \\
                    & T + P            & T+ P+ lens  &                    & T + P                     & T + P + lens       \\ 
 \hline \hline
 $h$                           &4.21\%  &2.71\%    &58.42\%    &0.99\%       & 0.92\%    \\ 
 $h^2 \Omega_b$  &0.56\%  &0.52\%    &164.1\%  &0.52\%      &0.52\%        \\ 
 $h^2 \Omega_c$  &1.03\%   &0.83\%   &130.96\%  & 0.87\%    &0.74\%       \\
 $n_s$                      &0.81\%   &0.45\%   & 70\%   &0.21\%     &0.16\%        \\
 $\alpha$                 &1.83\%   &1.48\%    &not constrained  &1.0\%    & 0.93\%          \\
 $\beta$                   & -             & -              &-              &-                 &-                    \\
 w                             &7.1\%      & 4.8\%      &not constrained   &1.2\%        &1.2\%           \\

\hline \hline
\end{tabular}
\end{center}
\end{table*}

\begin{table*}
\begin{center}
\caption{The same as Table \ref{tbl-modelo1}, but in this case $\beta=0$ will be kept fixed. \label{tbl-modelo1_sem_beta} }
\begin{tabular}{cccccc}
\hline
Parameter & CMB alone & CMB alone & P(k) alone &CMB + P(k)          & CMB + P(k)     \\
                    & Planck         & Planck         & SDSS        & Planck + SDSS   & Planck +SDSS \\
                    & T + P            & T+ P+ lens  &                    & T + P                     & T + P + lens       \\
 \hline \hline

 $h$                           &0.021     &0.019        &0.43     &0.0072        & 0.0066    \\ 
 $h^2 \Omega_b$  &0.00012 &0.00011    &0.038  &0.00011     &0.00011   \\ 
 $h^2 \Omega_c$  &0.0011    &0.00082    &0.13    & 0.00082     &0.00068   \\
 $n_s$                      &0.0039   &0.0036      & 0.38   &0.0013       &0.0013     \\
 $\alpha$                 &0.00094 &0.00084    &0.12    &0.000562    & 0.000558  \\
 w                             &0.045       &0.042        &0.91    &0.012           &0.012       \\
 \hline 
 \end{tabular} 

 \begin{tabular}{c}
          Percentage of the parameters' fiducial values for each error above  \\
   \end{tabular} 
   
  \begin{tabular}{cccccc}
  \hline
Parameter & CMB alone & CMB alone & P(k) alone &CMB + P(k)          & CMB + P(k)     \\
                    & Planck         & Planck         & SDSS        & Planck + SDSS   & Planck +SDSS \\
                    & T + P            & T+ P+ lens  &                    & T + P                     & T + P + lens       \\ 
 \hline \hline
 $h$                           &2.85\%    &2.58\%      &58.42\%    &0.98\%       & 0.90\%    \\ 
 $h^2 \Omega_b$  &0.52\%     &0.47\%     &164.1\%    &0.47\%       &0.47\%      \\ 
 $h^2 \Omega_c$  &1.03\%     &0.77\%     &121.61\%  & 0.77\%     &0.64\%        \\
 $n_s$                      &0.39\%    &0.36\%     & 38\%         &0.135\%       &0.129\%        \\
 $\alpha$                 &1.57\%      &1.40\%   &not constrained       &0.94\%      & 0.90\%        \\
 w                             &4.5\%       & 4.2\%      &91\%           &1.2\%        &1.2\%            \\

\hline \hline 
\end{tabular}
\end{center}
\end{table*}

We found the best upper limits for the isocurvature contribution from this  axion type of non adiabatic fluctuation, considering a $\Lambda$CDM cosmological model, for the combined Planck (considering CMB lensing) + SDSS  forecast with $\alpha < 0.061$ (95\% CL) and $ -0.0007<\beta <0.0007$ (95\% CL). If $\beta$ is not allowed to vary, the result for $\alpha$' s upper limit is not significantly changed. In this first scenario, we can also see in Figure \ref{fisher_modelo1} that lensing information can break the degeneracy between $\alpha$ and $\beta$.

\begin{figure*} [h!]
\includegraphics[scale=0.35]{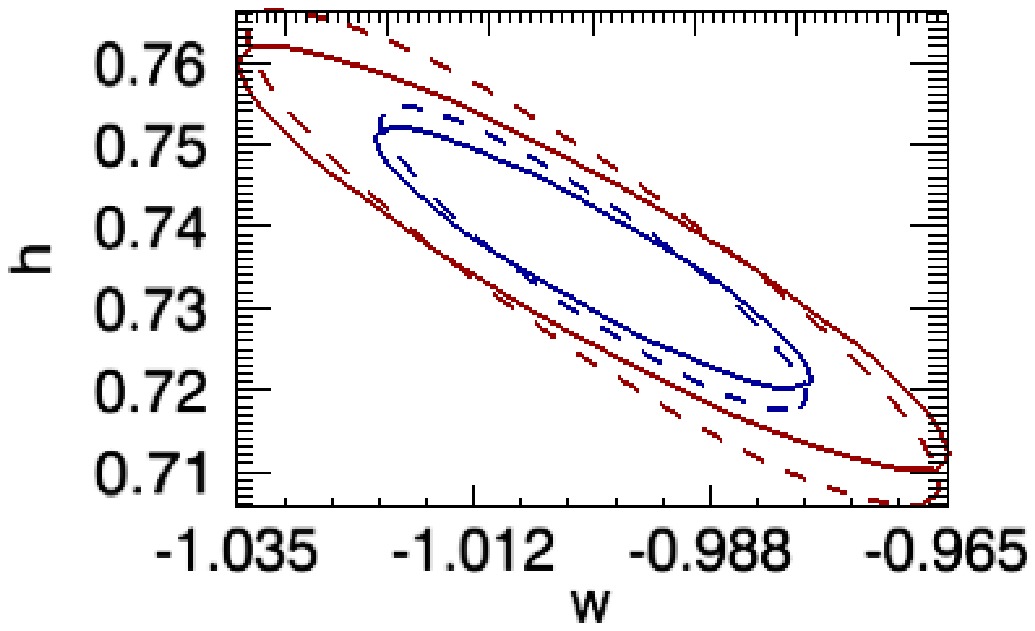}
\includegraphics[scale=0.35]{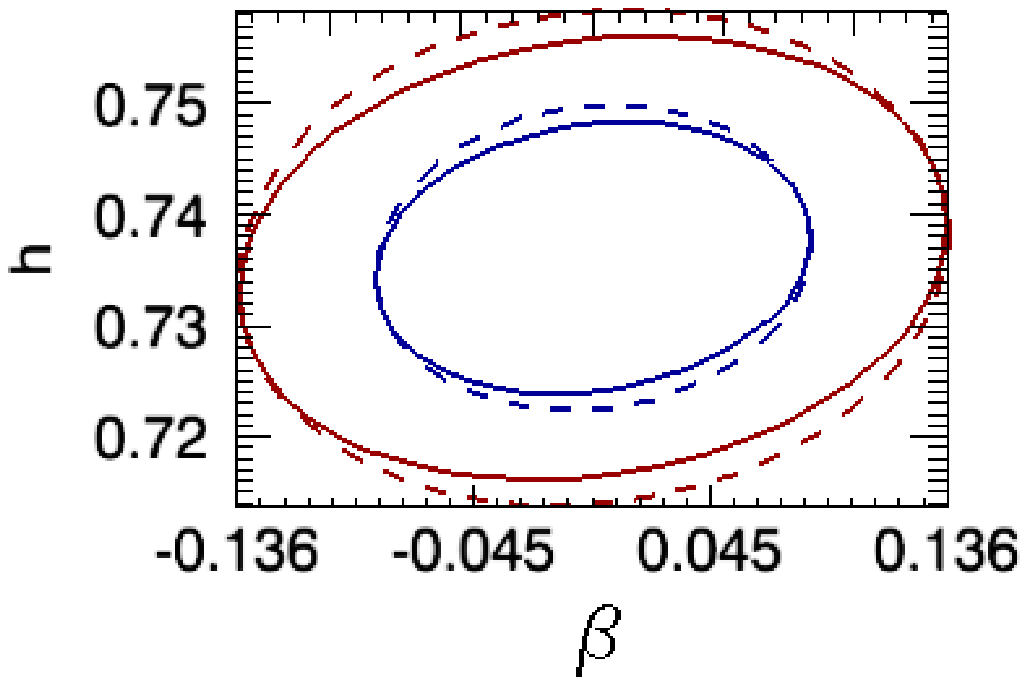} \\

\includegraphics[scale=0.35]{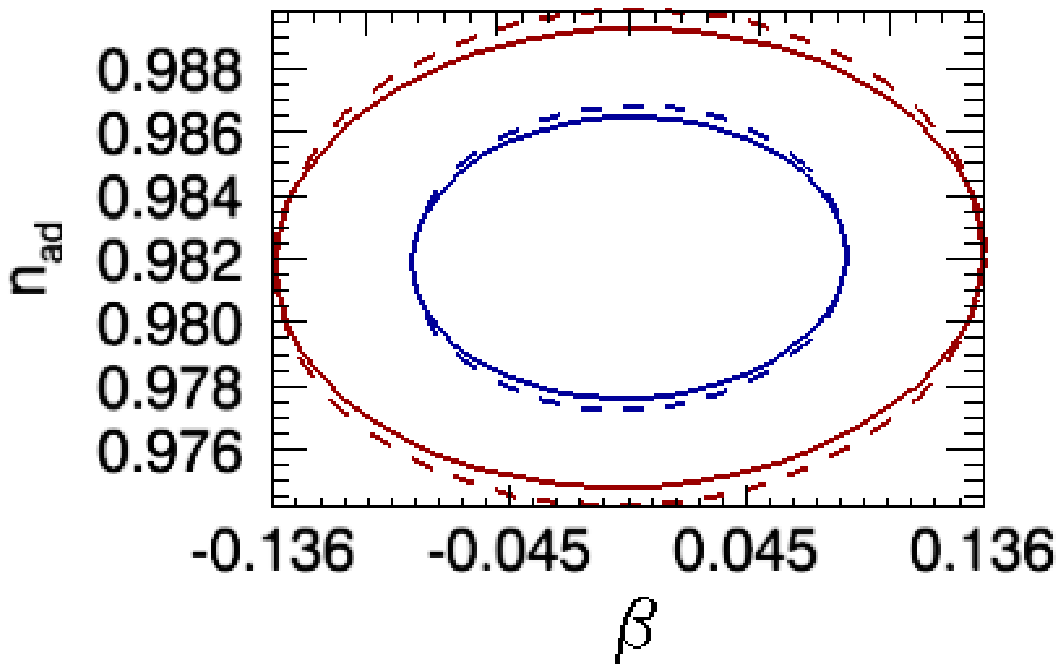} 
\includegraphics[scale=0.35]{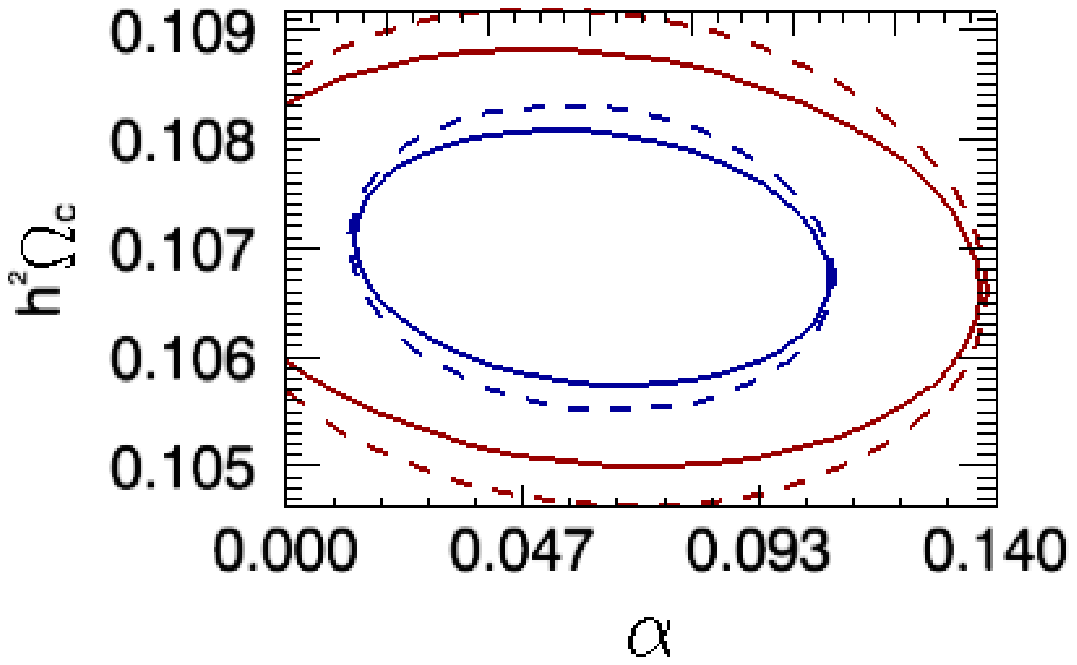}\\ 

\includegraphics[scale=0.35]{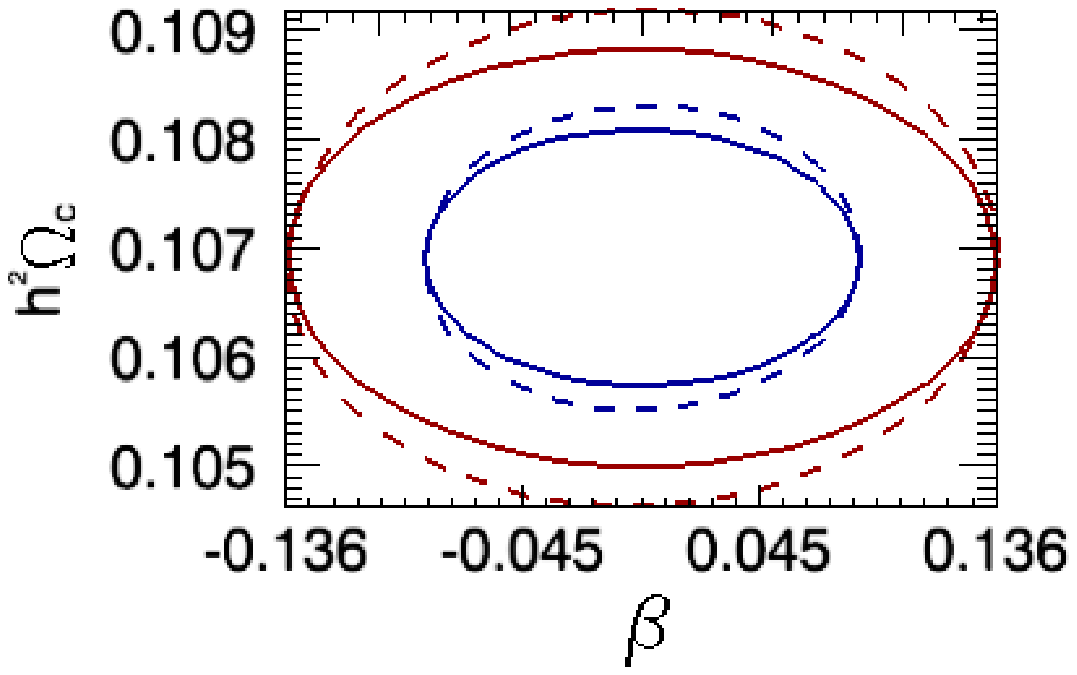} 
\includegraphics[scale=0.35]{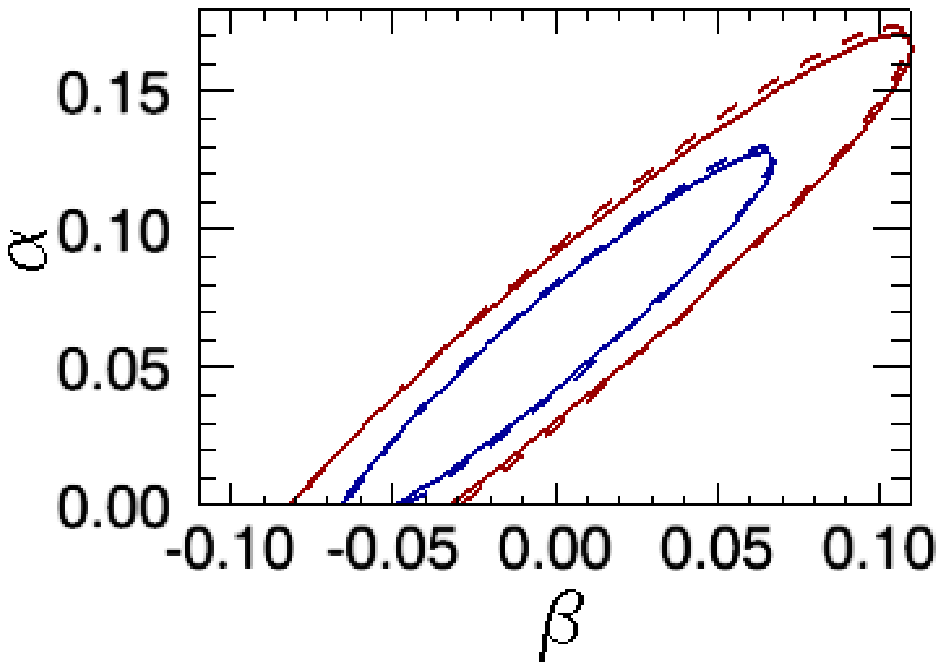} 

\caption{Fisher contours for  $\Lambda$CDM model plus a contribution of initial isocurvature fluctuation with fiducials amplitude of $\alpha = 0.06$, correlation phase of $\beta = 0$ and scalar spectral index of $n_{iso} = 0.982$ for the axion scenario. The red and blue contours represent 95.4\% and 68\% C.L. respectively for the unlesed CMB + SDSS (dashed lines) and for the lensed CMB + SDSS (solid lines) (see Table \ref {tbl-modelo3}).  } 
\label{fisher_modelo3}
\end{figure*}

On the other hand, for the second axion scenario, where the cosmological parameters' values are the same as the above ones, except for $n_{iso} = 0.982$ fixed  (this assumption was made by \citet{2006bean} and by the WMAP team following \citet{2009dunkley}), the isocurvature amplitude is better constrained when $\beta$ is kept fixed (see Tables \ref{tbl-modelo3}, \ref {tbl-modelo3_sem_beta} and Figure \ref{fisher_modelo3} for the Fisher constraints). The value for the limit of $\alpha$ for  the Planck forecast only (without including CMB lensing and keeping $\beta$ fixed) is comparable to the value found by the WMAP team: $\alpha < 0.11$  (95\% CL) for Planck, against $\alpha < 0.13$  (95\% CL) for WMAP 7-year data only \citep{2011larson}. However if we consider the CMB lensing in the analysis we can improve this limit, obtaining $\alpha < 0.10$ (95\% CL) for Planck. Finally, combining Planck (including CMB lensing) + SDSS, we have $\alpha < 0.08$ (95\% CL) against  $\alpha < 0.064$ (95\% CL)  found earlier with WMAP + BAO + SN \citep{2011komatsu}. If, on the other side, $\beta$ is allowed to vary we have that $\alpha < 0.12$ (95\% CL) as our best constraint (Planck + lensing information+ SDSS) and $-0.11<\beta<0.11$ (95\% CL). Even though worse constrained  we found an upper limit to $\alpha$ when $\beta$ is allowed to vary, giving us an extra bonus to constrain also the correlation phase. 

Since the values chosen for $n_{iso}$ in these first two scenarios are in the limit of the error  bars found using Lyman-$\alpha$ data, $n_{iso}=1.9 \pm 1$, for a sake of completeness we tested another scenario for $n_{iso}=1.9$ finding a significative change only in the isocurvature amplitude $\alpha$ ($\beta$ kept fixed). In this case, we found that  $\alpha < 0.066$ (95\% CL)  against $\alpha < 0.062$ (95\% CL)  for $n_{iso}=2.7$ and $\alpha < 0.1$ (95\% CL) for $n_{iso}=0.982$ considering for all them Planck only with CMB lensing information. As expected,  $\alpha$ is better constrained for higher  $n_{iso}$ values.

\begin{table*}
\begin{center}
\caption{Marginalized errors for $\Lambda$CDM model plus a contribution of initial isocurvature fluctuation with fiducials amplitude of $\alpha = 0.06$, correlation phase of $\beta = 0$ and scalar spectral index of $n_{iso} = n_{ad}=0.982$.\label{tbl-modelo3} }
\begin{tabular}{cccccc}
\hline
Parameter & CMB alone & CMB alone & P(k) alone &CMB + P(k)          & CMB + P(k)     \\
                    & Planck         & Planck         & SDSS        & Planck + SDSS   & Planck +SDSS \\
                    & T + P            & T+ P+ lens  &                    & T + P                     & T + P + lens       \\
 \hline \hline

 $h$                           &0.051         &0.036          &0.31       &0.0090        & 0.0081    \\ 
 $h^2 \Omega_b$  &0.00012      &0.00011    &0.028     &0.00011     &0.00011  \\ 
 $h^2 \Omega_c$  &0.0011         &0.00089    &0.095    & 0.00092    &0.00077   \\
 $n_s$                      &0.0033          &0.0030     & 0.26     &0.0031        &0.0029     \\
 $\alpha$                 &0.031           &0.031         &22.55    &0.031           & 0.031  \\
 $\beta$                   &0.088            &0.079         &41.78   &0.055          &0.054   \\
 w                             &0.13               &0.093         &5.40      &0.016         &0.015       \\
 \hline 
 \end{tabular} 

 \begin{tabular}{c}
          Percentage of the parameters' fiducial values for each error above  \\
   \end{tabular} 
   
  \begin{tabular}{cccccc}
  \hline
Parameter & CMB alone & CMB alone & P(k) alone &CMB + P(k)          & CMB + P(k)     \\
                    & Planck         & Planck         & SDSS        & Planck + SDSS   & Planck +SDSS \\
                    & T + P            & T+ P+ lens  &                    & T + P                     & T + P + lens       \\ 
 \hline \hline
 $h$                           &6.93\%  &4.89\%       &42.12\%    &1.22\%       & 1.10\%     \\ 
 $h^2 \Omega_b$  &0.52\%  &0.47\%       &120.95\%  &0.47\%       &0.47\%       \\ 
 $h^2 \Omega_c$  &1.03\%     &0.83\%       &88.87\%    &0.86\%       &0.72\%        \\
 $n_s$                      &0.33\%   &0.30\%      & 26\%         &0.31\%       &0.29\%         \\
 $\alpha$                 &52.48\%  &52.48\%   &not constrained   &52.48\%    & 51.48\%        \\
 $\beta$                & -             & -              &-              &-                 &- \\
 w                             &13\%         & 9.3\%      &not constrained         &1.6\%       &1.5\%            \\

\hline \hline 
\end{tabular}
\end{center}
\end{table*}

\begin{table*}
\begin{center}
\caption{The same as Table \ref{tbl-modelo3}, but in this case $\beta=0$ will be kept fixed. \label{tbl-modelo3_sem_beta} }
\begin{tabular}{cccccc}
\hline
Parameter & CMB alone & CMB alone & P(k) alone &CMB + P(k)          & CMB + P(k)     \\
                    & Planck         & Planck         & SDSS        & Planck + SDSS   & Planck +SDSS \\
                    & T + P            & T+ P+ lens  &                    & T + P                     & T + P + lens       \\
 \hline \hline

 $h$                           &0.032     &0.024          &0.30      &0.0090         & 0.0080    \\ 
 $h^2 \Omega_b$  &0.00012 &0.00011     &0.028    &0.00011       &0.00011   \\ 
 $h^2 \Omega_c$  &0.0010   &0.00086      &0.090    & 0.00092     &0.00077   \\
 $n_s$                      &0.0034   &0.0030       & 0.26      &0.0031         &0.0029     \\
 $\alpha$                 &0.025     &0.020       &2.62           &0.010           & 0.0099     \\
 w                             &0.080      &0.063         &5.39         &0.016           &0.015       \\
 \hline
 \end{tabular} 

 \begin{tabular}{c}
          Percentage of the parameters' fiducial values for each error above  \\
   \end{tabular} 
   
  \begin{tabular}{cccccc}
  \hline
Parameter & CMB alone & CMB alone & P(k) alone &CMB + P(k)          & CMB + P(k)     \\
                    & Planck         & Planck         & SDSS        & Planck + SDSS   & Planck +SDSS \\
                    & T + P            & T+ P+ lens  &                    & T + P                     & T + P + lens       \\ 
 \hline \hline
  $h$                           &4.35\%    &3.26\%      &40.76\%      &1.22\%       & 1.09\%    \\ 
 $h^2 \Omega_b$  &0.52\%    &0.47\%      &120.95\%    &0.47\%       &0.47\%      \\ 
 $h^2 \Omega_c$  &0.93\%       &0.80\%     &84.19\%       & 0.86\%      &0.72\%        \\
 $n_s$                      &0.34\%    &0.30\%     & 26\%            &0.31\%       &0.29\%        \\
 $\alpha$                 &41.67\%    &33.33\%   &not constrained     &17.5\%          & 16.65\%        \\
 w                             &8.0\%        & 6.3\%      &not constrained          &1.6\%         &1.5\%          \\

\hline \hline 
\end{tabular}
\end{center}
\end{table*}

\begin{figure*} 
\includegraphics[scale=0.35]{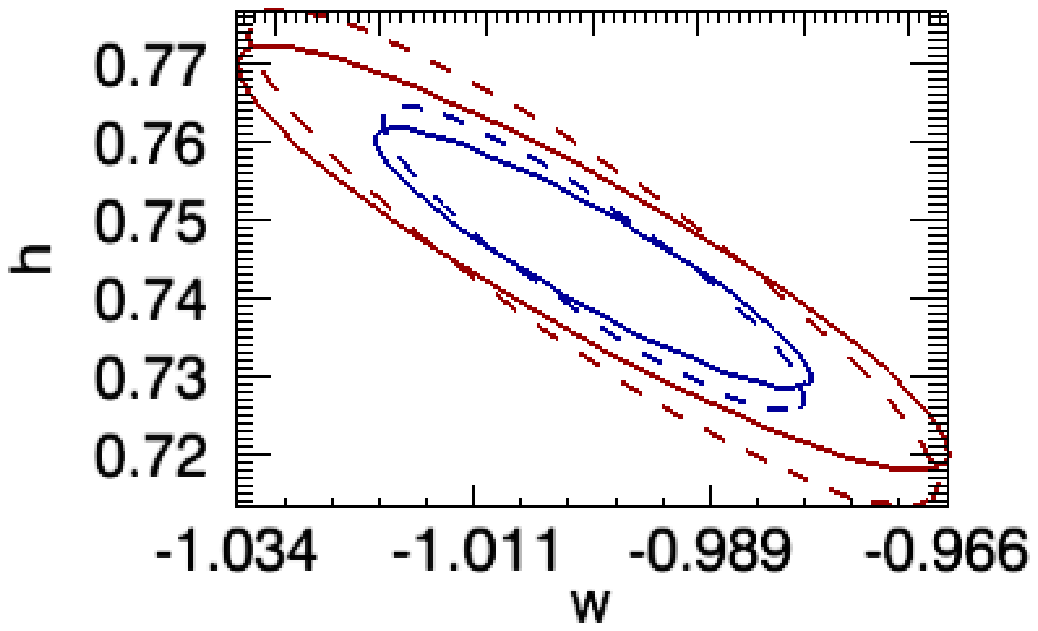}
\includegraphics[scale=0.35]{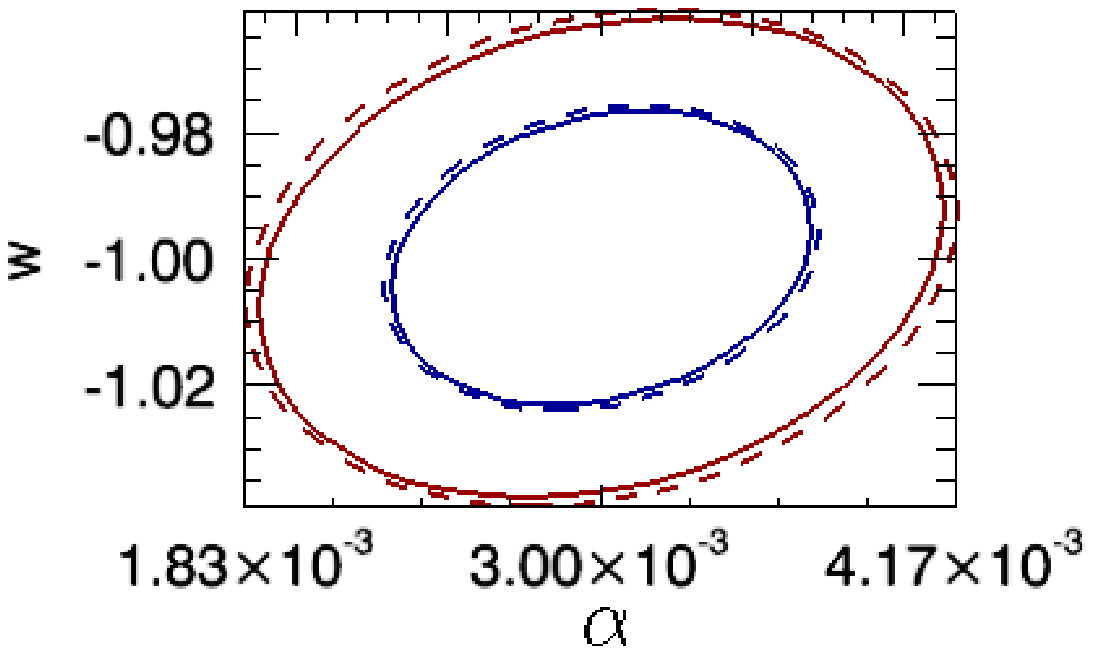} \\

\includegraphics[scale=0.35]{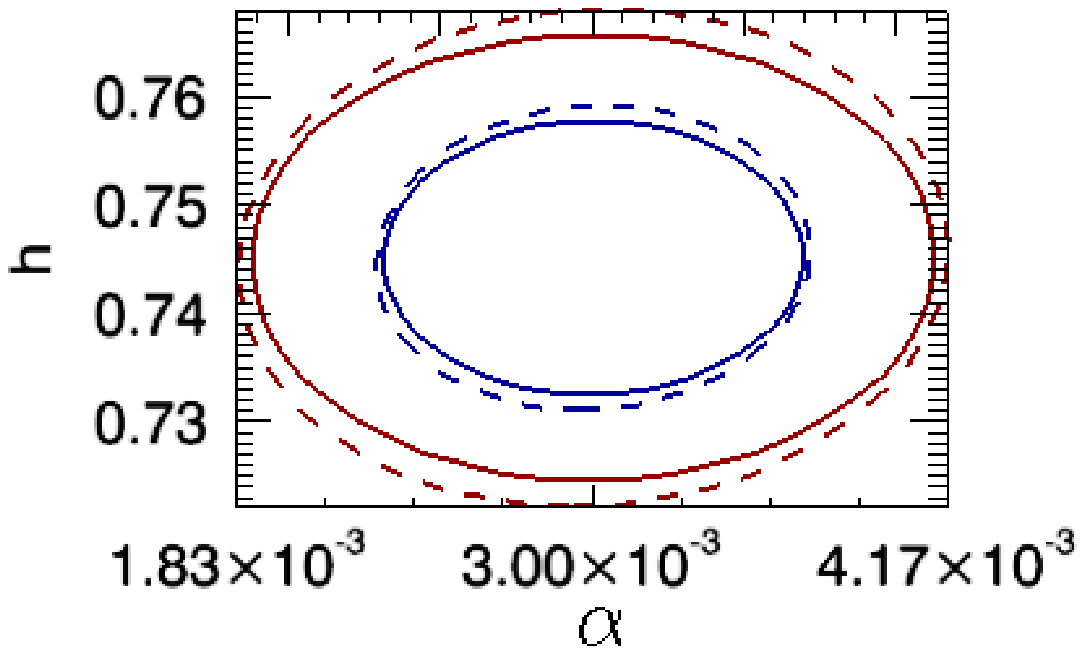} 
\includegraphics[scale=0.35]{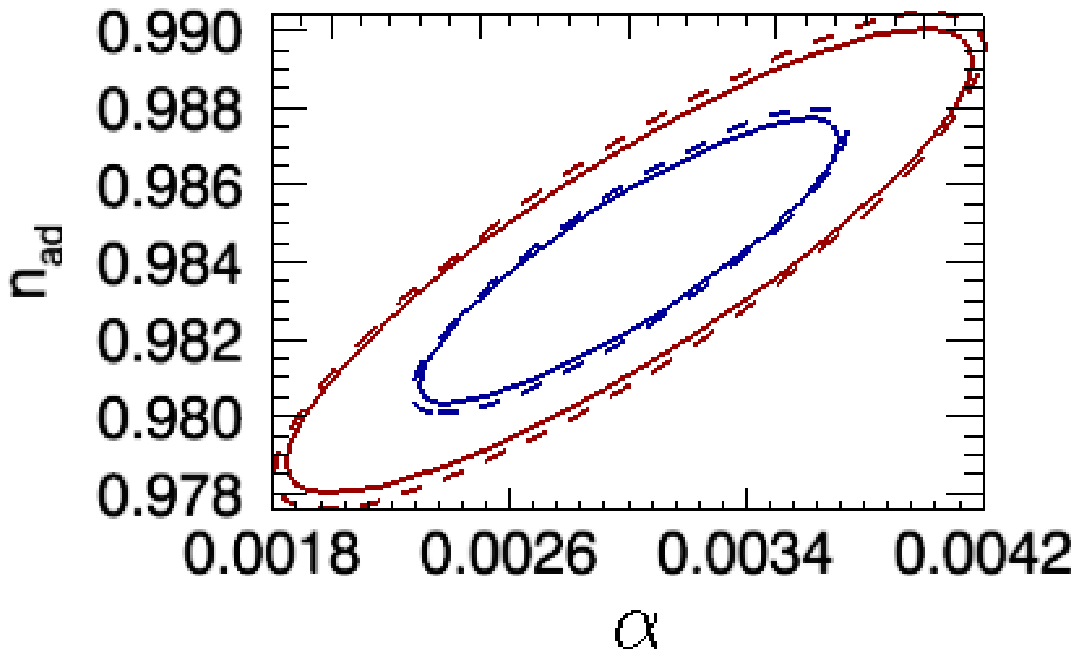}\\

\includegraphics[scale=0.35]{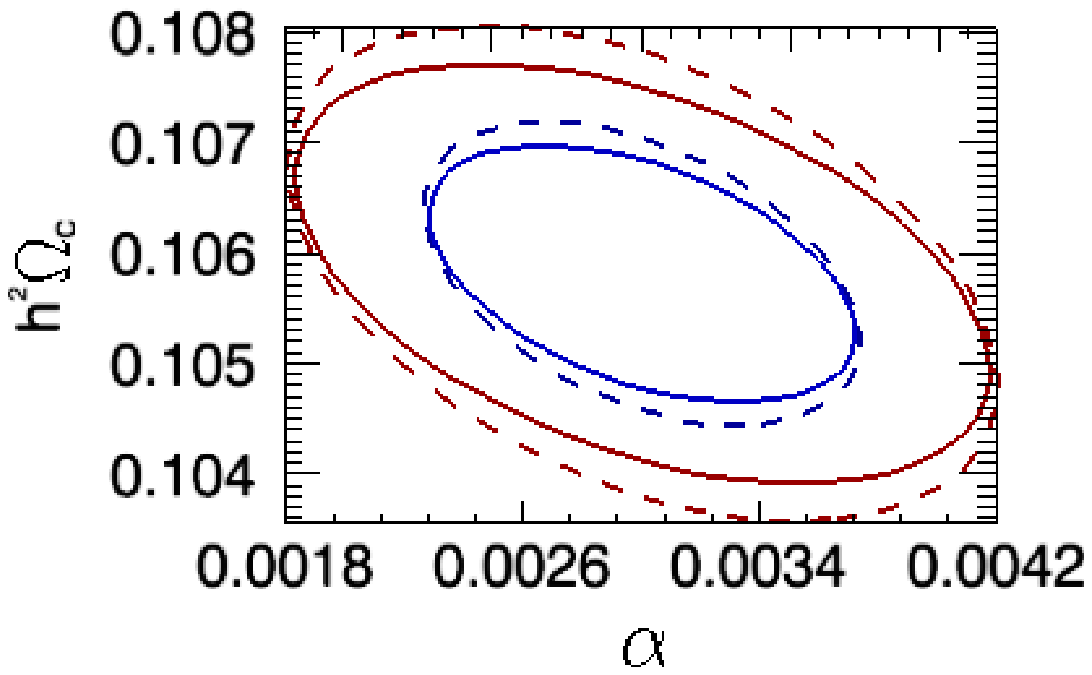}
\includegraphics[scale=0.35]{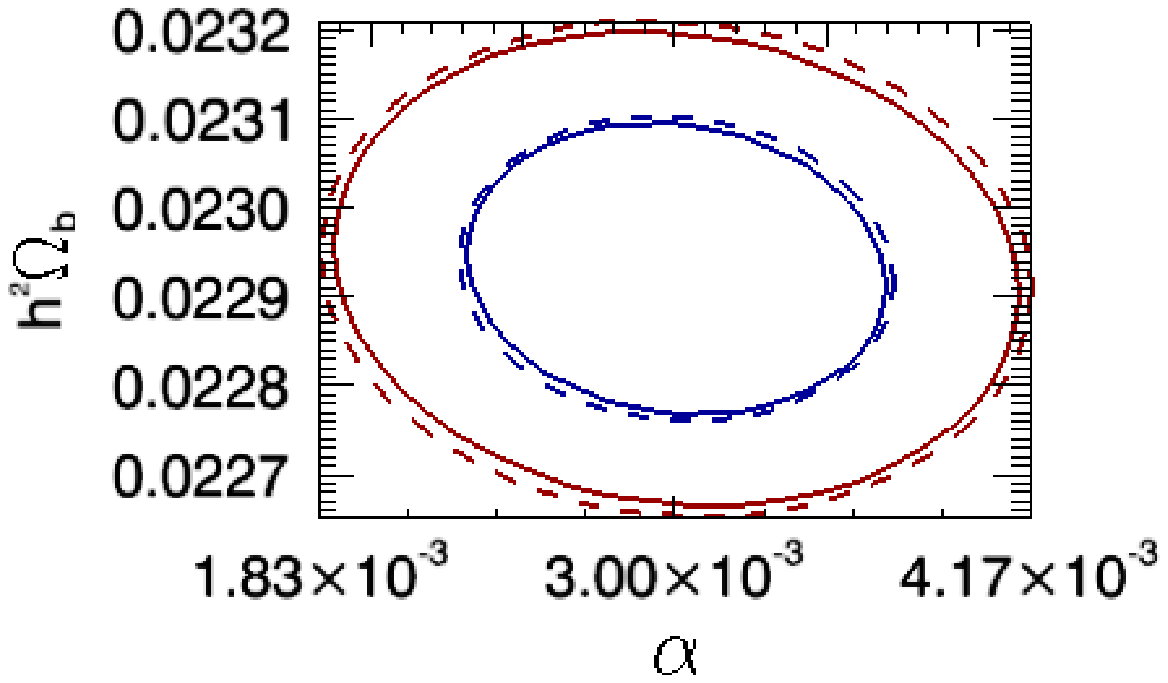}\\

\caption{Fisher contours for  $\Lambda$CDM model plus a contribution of initial isocurvature fluctuation with fiducials amplitude of $\alpha = 0.003$, correlation phase of $\beta = -1$ and scalar spectral index of $n_{iso} = 0.984$ for the curvaton scenario. The red and blue contours represent 95.4\% and 68\% C.L. respectively for the unlesed CMB + SDSS (dashed lines) and for the lensed CMB + SDSS (solid lines). In this case we consider $\beta$ fixed (see Table \ref {tbl-modelo2_sem_beta}).}  
\label{fisher_modelo2}  
\end{figure*}

In the last scenario, the isocurvature primordial fluctuations are generated by the decay of the curvaton.  Our fiducial parameters' values are set to be, $h=0.745$, $\Omega_b =0.02293 h^2$, $\Omega_c=0.1058 h^2$, $n_{ad} = 0.984$ ($n_{iso} = 0.984$ fixed as generally predicted by curvaton scenarios \citep{2006bean}), $\alpha =0.003$, $\beta= -1$ and $w=-1$.  Unlike the first two cases, none of the isocurvature parameters can be constrained if $\beta$ is allowed to vary, as it can be seen in Table \ref{tbl-modelo2}. Nevertheless, the upper limit found for $\alpha$ ($\beta$ fixed) is improved, with Planck only, $\alpha < 0.0068$ (95\% CL), compared to the one found for WMAP 7-year data only, $\alpha < 0.011$ (95\% CL). An even better constraint is reached considering the CMB lensing effect in our analysis (see, Table \ref{tbl-modelo2_sem_beta} and Figure \ref{fisher_modelo2}), improving Planck limit to $\alpha < 0.0054$ (95\% CL).  For Planck (including CMB lensing) + SDSS, $\alpha < 0.0039$ (95\% CL) against $\alpha < 0.0037$ (95\% CL) for WMAP + BAO + SN. Using therefore other cosmological probes, such as BAO and SN to Planck with lensing information + SDSS, the error bars in the isocurvature amplitude can become even smaller,  allowing to a very limited isocurvature contribution in the primordial fluctuations when the it is completely anti-correlated with the adiabatic component.

\begin{table*}
\begin{center}
\caption{Marginalized errors for $\Lambda$CDM model plus a contribution of initial isocurvature fluctuation with fiducials amplitude of $\alpha = 0.003$, correlation phase of $\beta = -1$ and scalar spectral index of $n_{iso} = 0.984$.\label{tbl-modelo2} }
\begin{tabular}{cccccc}
\hline
Parameter & CMB alone & CMB alone & P(k) alone &CMB + P(k)          & CMB + P(k)     \\
                    & Planck         & Planck         & SDSS        & Planck + SDSS   & Planck +SDSS \\
                    & T + P            & T+ P+ lens  &                    & T + P                     & T + P + lens       \\
 \hline \hline

 $h$                           &0.055     &0.036        &0.33        &0.0094        & 0.0084    \\ 
 $h^2 \Omega_b$  &0.00012 &0.00011   &0.029      &0.00011     &0.00011   \\ 
 $h^2 \Omega_c$  &0.0011   &0.00090    &0.096     & 0.00093     &0.00079   \\
 $n_s$                      &0.0030   &0.0028      & 0.26      &0.0029        &0.0027     \\
 $\alpha$                 &0.029     &0.030         &17.63     &0.028          & 0.029      \\
 $\beta$                   &4.98        &5.19           &3084.60 &4.95           &5.10        \\
 w                             &0.13        &0.089          &7.48       &0.016         &0.015      \\
 \hline 
 \end{tabular} 

 \begin{tabular}{c}
          Percentage of the parameters' fiducial values for each error above  \\
   \end{tabular} 
   
  \begin{tabular}{cccccc}
  \hline
Parameter & CMB alone & CMB alone & P(k) alone &CMB + P(k)          & CMB + P(k)     \\
                    & Planck         & Planck         & SDSS        & Planck + SDSS   & Planck +SDSS \\
                    & T + P            & T+ P+ lens  &                    & T + P                     & T + P + lens       \\ 
 \hline \hline
 $h$                           &7.38\%  &4.83\%          &44.29\%                           &1.26\%       & 1.13\%    \\ 
 $h^2 \Omega_b$  &0.52\%     &0.48\%          &126.47\%                         &0.48\%       &0.48\%      \\ 
 $h^2 \Omega_c$  &1.04\%     &0.85\%        &90.73\%                           & 0.88\%      &0.75\%       \\
 $n_s$                      &0.30\%     &0.28\%        & 26\%                               &0.29\%        &0.27\%        \\
 $\alpha$                 &not constrained      &not constrained       &not constrained  & not constrained       & not constrained   \\
 $\beta$\                   &not constrained       & not constrained      & not constrained   &495\%          & not constrained        \\
 w                             &13\%          & 8.9\%         &not constrained                              &1.6\%            &1.5\%           \\

\hline \hline 
\end{tabular}
\end{center}
\end{table*}

\begin{table*}
\begin{center}
\caption{The same as Table \ref{tbl-modelo2}, but in this case $\beta=-1$ will be kept fixed. \label{tbl-modelo2_sem_beta} }
\begin{tabular}{cccccc}
\hline
Parameter & CMB alone & CMB alone & P(k) alone &CMB + P(k)          & CMB + P(k)     \\
                    & Planck         & Planck         & SDSS        & Planck + SDSS   & Planck +SDSS \\
                    & T + P            & T+ P+ lens  &                    & T + P                     & T + P + lens       \\
 \hline \hline

 $h$                           &0.055      &0.0345         &0.31      &0.0092         & 0.0083    \\ 
 $h^2 \Omega_b$  &0.00012  &0.00011      &0.027    &0.00011      &0.00011   \\ 
 $h^2 \Omega_c$  &0.0010     &0.00088      &0.092    & 0.00091     &0.00077   \\
 $n_s$                      &0.0027     &0.0025       &0.24      &0.0026         &0.0024     \\
 $\alpha$                 &0.0019     &0.0012       &0.17      &0.00047       & 0.00045  \\
 w                             &0.14           &0.088          &7.47      &0.016          &0.015       \\
 \hline 
 \end{tabular} 

 \begin{tabular}{c}
          Percentage of the parameters' fiducial values for each error above  \\
   \end{tabular} 
   
  \begin{tabular}{cccccc}
  \hline
Parameter & CMB alone & CMB alone & P(k) alone &CMB + P(k)          & CMB + P(k)     \\
                    & Planck         & Planck         & SDSS        & Planck + SDSS   & Planck +SDSS \\
                    & T + P            & T+ P+ lens  &                    & T + P                     & T + P + lens       \\ 
 \hline \hline
 $h$                           &7.38\%    &4.63\%        &41.61\%       &1.23\%       & 1.11\%    \\ 
 $h^2 \Omega_b$  &0.52\%       &0.48\%      &117.75\%    &0.48\%       &0.48\%      \\ 
 $h^2 \Omega_c$  &0.94\%       &0.83\%      &87.90\%       & 0.86\%      &0.72\%        \\
 $n_s$                      &0.27\%       &0.25\%      & 24\%           &0.26\%       &0.24\%        \\
 $\alpha$                 &63.33\%     &40\%    &not constrained  &15.67\%         & 15\%        \\
 w                             &14\%           & 8.8\%          &not constrained         &1.6\%        &1.5\%            \\

\hline \hline
\end{tabular}
\end{center}
\end{table*}

\section{Discussion and conclusions\label{Discussion and conclusions}}

In this paper we studied a possible contribution of isocurvature initial perturbations in the pure adiabatic fluctuations scenario from the well tested $\Lambda$CDM model. Using the fisher formalism we obtained the best constraints possible for the isocurvature parameters using CMB and galaxy distribution information.  

The main goal of this work has been to quantify how CMB lensing information can provide better constraints in the cosmological parameters, specially in the ones related to the isocurvature contribution. Moreover, we saw that CMB lensing information broke the parameter degeneracie between the isocurvature parameters $\alpha$ and $\beta$ for one of the three studied scenarios. 

In all  tested inflationary scenarios, the CMB lensing information improves the constraints of all chosen parameters, including the ones related to the isocurvature mode.  If we consider Planck information alone (with $\beta$ not allowed to vary) the smallest improvement obtained  on $\alpha$  standard deviation is in the axion type inflation for $n_{ad}\neq n_{iso}$ with a difference of 0.17\% of its fiducial value between the lensed and unlensed analysis. This improvement gets bigger for the scenarios considered by WMAP reaching almost 9\% (axion type with $n_{ad} = n_{iso}$) and 25\% (curvaton type) (see the lower part  where  $\beta$ is kept fixed inTables \ref{tbl-modelo1_sem_beta}, \ref{tbl-modelo3_sem_beta} and \ref{tbl-modelo2_sem_beta}).  

\begin{figure*} [h]
\includegraphics[scale=0.6]{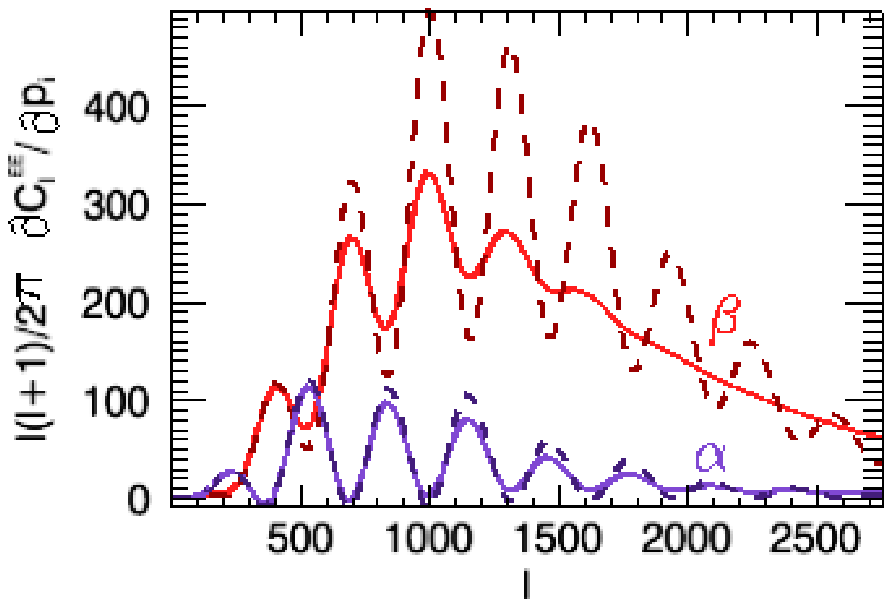} 
\includegraphics[scale=0.6]{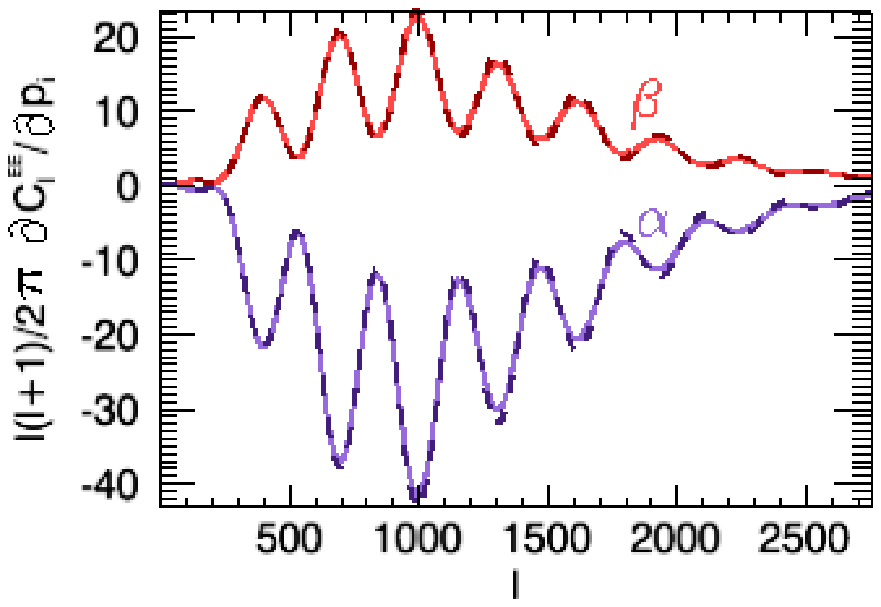} \\

\includegraphics[scale=0.6]{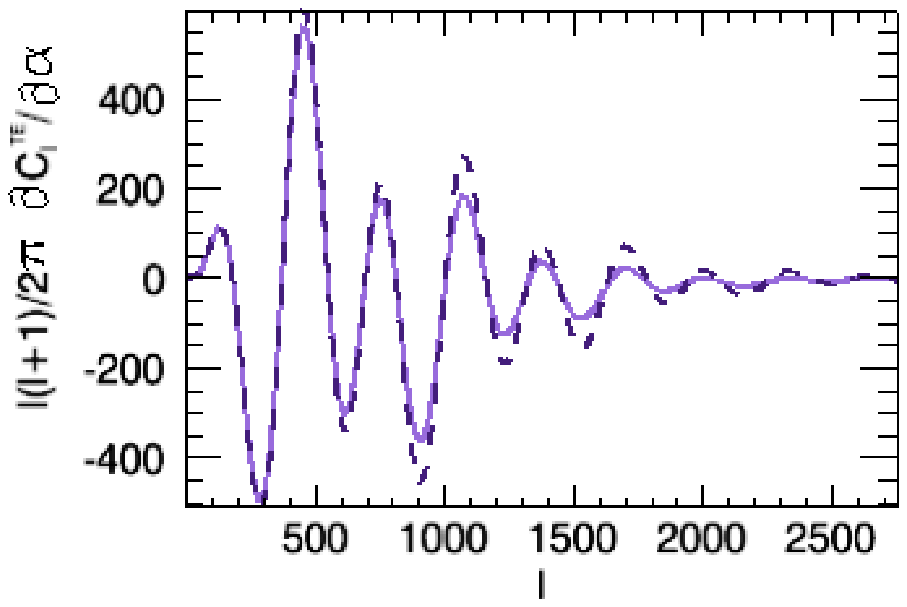} 
\includegraphics[scale=0.6]{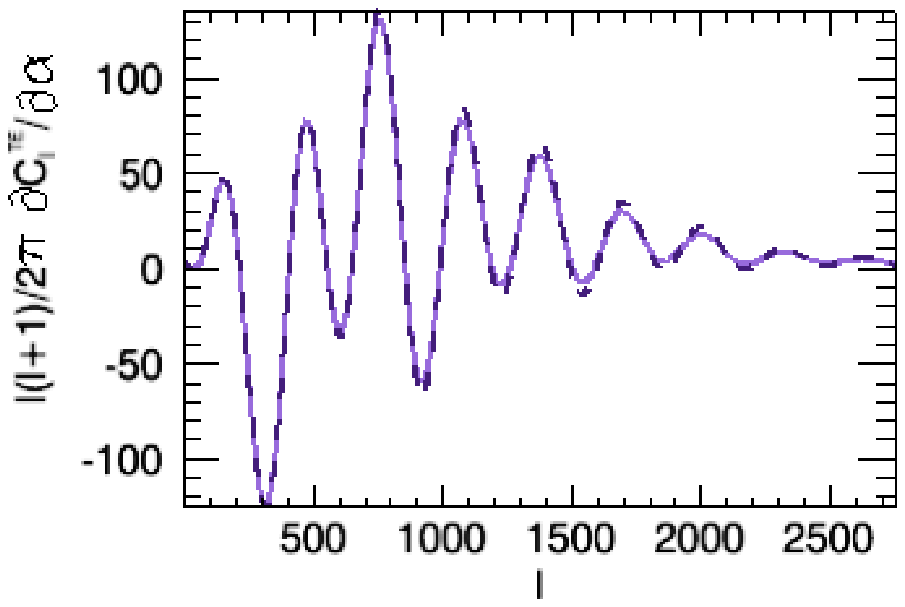} \\

\includegraphics[scale=0.6]{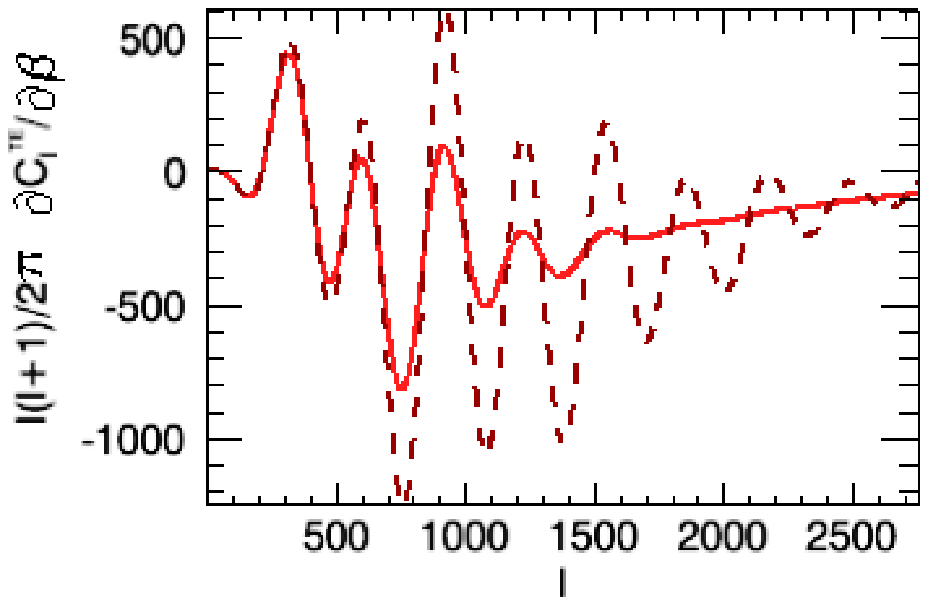}
\includegraphics[scale=0.6]{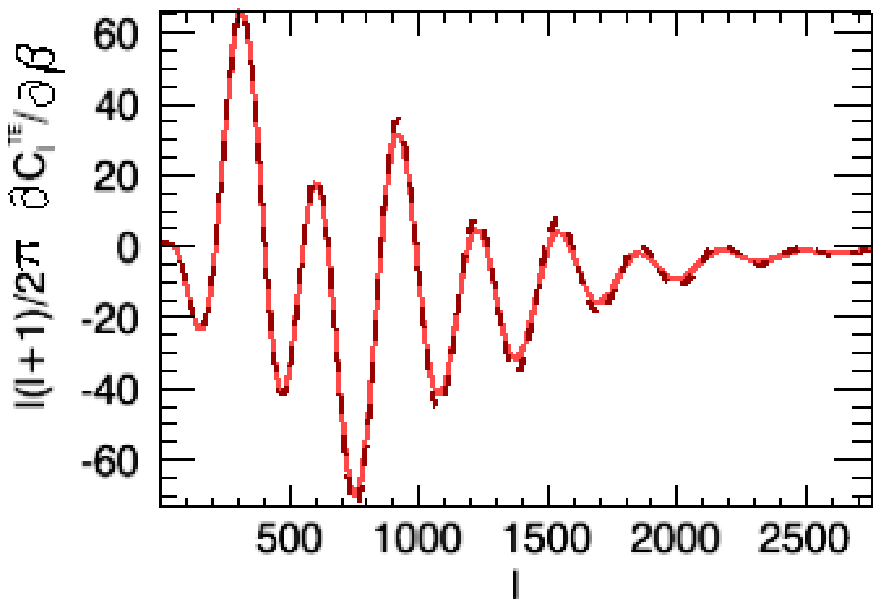}

\caption{ CMB power spectra derivatives in respect to the isocurvature parameters $\alpha$ (purple) and $\beta$(red). The dashed darker lines are related to the unlensed power spectra's derivative and the solid lighter ones are related to the lensed power spectra's derivative for the axion scenario of $\alpha = 0.06$, $\beta = 0$. On the left column the scalar spectral index is $n_{iso} = 2.7$ and on the right column the scalar spectral index is $n_{iso} = 0.982$ }
\label{der1} 
\end{figure*}

Moreover, if CMB lensing can be measured, it would be possible to distinguish between the axion models with $n_{iso}= 0.982$ and $n_{iso}= 2.7$ for instance. The effect of CMB lensing is bigger for higher $n_{iso}$ values as can be seen in the comparison of Figures \ref{fisher_modelo1} and \ref{fisher_modelo3}. We can visualize better this lensing effect on  $n_{iso}$  by analyzing the power spectra derivatives in respect to $\alpha$ and $\beta$ in Figure \ref{der1}.

When the combined Planck + SDSS forecast is done, the improvement with the use of lensing information is not so significant  for any of the scenarios. This is due to the poor ability of SDSS to constrain the parameters compared to Planck, especially when CMB lensing information is included.  For a CMB experiment alone, or combined with any other precise experiments on galaxies distribution,  lensing is  an important extra information in the attempt to know how well observations can constrain the presence  of isocurvature contribution to the primordial fluctuations. An interesting forecast would include future galaxy surveys, such as EUCLID, combined with planck CMB information including the lensing effects. 

\appendix

\section{Elements of the covariance matrix and lensing corrections}

The elements of the covariance matrix in the unlensed case are:

\begin{equation}
\label {cov1}
\Xi_{TTTT} = (C^{TT}_l + N^{TT}_l)^2,
\end{equation}

\begin{equation}
\label {cov2}
\Xi_{EEEE} =  (C^{EE}_l + N^{PP}_l)^2,
\end{equation}

\begin{equation}
\label {cov3}
\Xi_{BBBB} =  (C^{BB}_l + N^{PP}_l)^2,
\end{equation}

\begin{equation}
\label {cov4}
\begin{split}
\Xi_{TETE} =& (C^{TE}_l )^2+ (C^{TT}_l + N^{TT}_l) \\
 &\times (C^{EE}_l + N^{PP}_l),
\end{split}
\end{equation}

 \begin{equation}
 \label {cov5}
\Xi_{TTEE} =  (C^{TE}_l )^2,
\end{equation}

\begin{equation}
\label {cov6}
\Xi_{TTTE} = C^{TE}_l  (C^{TT}_l + N^{TT}_l),
\end{equation}

\begin{equation}
\label {cov7}
\Xi_{EETE} = C^{TE}_l  (C^{EE}_l + N^{PP}_l),
\end{equation}

\begin{equation}
\label {cov8}
\Xi_{TTBB} = \Xi_{EEBB} = \Xi_{TEBB} = 0.
\end{equation}

In these equations,  $N^{TT}_l$ and $N^{PP}_l$ are the gaussian random detector noises for  temperature and polarization respectively, which expression is written using the window function, $B_l^2 = exp[-l(l+1)\theta_{beam}^2 / 8\ln2]$ and the inverse square of the detector noise level for temperature and polarization, $w_T$ and $w_P$.  The Full Width Half Maximum (FWHM), $\theta_{beam}$, is used in radians and $w=(\theta_{beam}\sigma)^{-2}$ is the weight given to each considered  Planck channel  \citep{1999eisenstein} . The experimental specifications can be checked in Table \ref {planck}.

\small
\begin{equation}
\label {noise1}
N^{TT}_l = [(w_TB_l^{2})_{100}+ (w_TB_l^{2})_{143}+ (w_TB_l^{2})_{217}+ (w_TB_l^{2})_{353}]^{-1}
\end{equation}

\begin{equation}
\label {noise2}
N^{PP}_l = [(w_PB_l^{2})_{100}+(w_PB_l^{2})_{143}+ (w_PB_l^{2})_{217}+  (w_PB_l^{2})_{353}]^{-1}
\end{equation}
\normalsize

Here we used four channels, 100, 143, 217 and 353GHz of the Planck experiment as can be seen from the equations \ref{noise1} and \ref{noise2}.

\begin{table}[h!]
\caption{Planck  \textrm{specifications\footnote{See Planck mission blue book at http://www.rssd.esa.int/SA/PLANCK/docs/Bluebook-ESA-SCI(2005)1\_V2.pdf} }\label{planck} }
\begin{ruledtabular}
\begin{tabular}{cccc}
Frequency (GHz) & $\theta_{beam}$ & $\sigma_T(\mu K-arc)$ & $\sigma_P(\mu K-arc)$\\
 100 & 9.5' & 6.82        & 10.9120\\
 143  & 7.1' & 6.0016   & 11.4576\\ 
  217 & 5.0' & 13.0944 & 26.7644\\ 
  353 & 5.0' &40.1016  &81.2944\\
 \end{tabular}
 \end{ruledtabular}
\end{table}

Corrections for the lensed case are \citep{2006perotto}:

\small
\begin{equation}
\label {cov_l1}
\begin{split}
\xi_{TTTT} = &\left(\textrm{\~C}^{TT}_l + N^{TT}_l\right)^2\\
& - \frac{2\left(\textrm{\~C}^{TE}_l\right)^2\left(\textrm{\~C}^{Td}_l\right)^2}{\left(\textrm{\~C}^{EE}_l + N^{PP}_l\right)^2\left(C^{dd}_l + N^{dd}_l\right)^2},
\end{split}
\end{equation}
\normalsize

\begin{equation}
\label {cov_l2}
\xi_{EEEE} =  \left(\textrm{\~C}^{EE}_l + N^{PP}_l\right)^2,
\end{equation}

\begin{equation}
\label {cov_l3}
\xi_{BBBB} =  \left(\textrm{\~C}^{BB}_l + N^{PP}_l\right)^2,
\end{equation}

\small
\begin{equation}
\label {cov_l4}
\begin{split}
\xi_{TETE} = & \frac{1}{2} \left[\left(\textrm{\~C}^{TE}_l \right)^2+ \left(\textrm{\~C}^{TT}_l + N^{TT}_l\right) \left(\textrm{\~C}^{EE}_l + N^{PP}_l\right)\right] \\
&- \frac{\left(\textrm{\~C}^{EE}_l + N^{PP}_l\right)\left(C^{Td}_l\right)^2}{2\left(C^{dd}_l + N^{dd}_l\right)} ,
 \end{split}
\end{equation}
\normalsize

\small
\begin{equation}
\label {cov_l5}
\begin{split}
\xi_{TdTd} = & \frac{1}{2} \left[\left(C^{Td}_l \right)^2+ \left(\textrm{\~C}^{TT}_l + N^{TT}_l\right) \left(C^{dd}_l + N^{dd}_l\right)\right] \\
& - \frac{\left(C^{dd}_l + N^{dd}_l\right)\left(\textrm{\~C}^{TE}_l\right)^2}{2\left(\textrm{\~C}^{EE}_l + N^{EE}_l\right)} ,
 \end{split}
\end{equation}
\normalsize

 \begin{equation}
 \label {cov_l6}
\xi_{dddd} =  \left(C^{dd}_l + N^{dd}_l \right)^2,
\end{equation}

 \begin{equation}
 \label {cov_l7}
\xi_{TTEE} =  \left(\textrm{\~C}^{TE}_l \right)^2,
\end{equation}

\small
\begin{equation}
\label {cov_l8}
\xi_{TTTE} = \textrm{\~C}^{TE}_l \left[\left(\textrm{\~C}^{TT}_l + N^{TT}_l\right) - \frac{\left(C^{Td}_l\right)^2}{\left(C^{dd}_l + N^{dd}_l\right)}\right],
\end{equation}
\normalsize

\begin{equation}
\label {cov_l9}
\xi_{TEEE} = \textrm{\~C}^{TE}_l  \left(\textrm{\~C}^{EE}_l + N^{PP}_l\right),
\end{equation}

\begin{equation}
\label {cov_l10}
\xi_{TTdd} = \left(C^{Td}_l \right)^2,
\end{equation}

\small
\begin{equation}
\label {cov_l11}
\xi_{TTTd} = C^{Td}_l \left[(\textrm{\~C}^{TT}_l + N^{TT}_l) - \frac{(\textrm{\~C}^{TE}_l)^2}{(\textrm{\~C}^{EE}_l + N^{EE}_l)}\right],
\end{equation}
\normalsize

\begin{equation}
\label {cov_l12}
\xi_{Tddd} = C^{Td}_l \left(C^{dd}_l + N^{dd}_l \right),
\end{equation}

where $N^{dd}_l$ is the optimal quadratic estimator  \footnotetext[3]{Here we consider the TT quadratic estimator since it provides the best estimator for the Planck experiment (for a review, see \citet{2003okamoto, 2002hu})} noise of the deflection field and it can be written in the form \citep{2002bhu}:

\begin{equation}
\label {noise3}
N^{dd}_l= \left[ \sum_{l1l2}  \frac{(C_{l2}^{TT}F_{l1ll2} + C_{l1}^{TT}F_{l2ll1})^2}{2(\textrm{\~C}_{l1}^{TT} + N_{l1}^{TT})(\textrm{\~C}_{l2}^{TT} + N_{l2}^{TT})}\right]^{-1} ,
\end{equation}

\begin{equation}
\begin{split}
F_{l1ll2}= \sqrt{ \frac{(2l_1+1)(2l+1)(2l_2 +1)}{4\pi}}  \left(\begin{array}{rrr}
l_1 & l & l_2\\
0 & 0 & _0
\end{array}\right) \\
\times  \frac{1}{2} \left[l(l+1) +l_2(l_2+1)-l_1(l_1+1)\right]. \nonumber
\end{split}
\end{equation}

Note that the Fisher matrix analysis approximates the likelihood as Gaussian function, however the likelihood function could in general be non-Gaussian. Nonetheless, as stated in \citep{2006perotto} CMB lensing information gives a more Gaussian likelihood function, breaking some parameters degeneracies, consequently providing a better error estimation.

\end{document}